\documentclass[prl, aps, longbibliography, twocolumn, reprint, showpacs, footnoteinbib, superscriptaddress]{revtex4-1}
\usepackage{graphicx}
\usepackage{amssymb}   
\usepackage{amsfonts}
\usepackage{amsmath}
\usepackage{dsfont}
\usepackage{dcolumn}   
\usepackage{bm}        
\usepackage[mathscr]{eucal}
\usepackage[dvipsnames]{xcolor}
\usepackage[colorlinks, linkcolor=WildStrawberry,citecolor=WildStrawberry,urlcolor=PineGreen]{hyperref}
\usepackage[all]{hypcap} 
\usepackage{mathtools}
\usepackage{wasysym}
\usepackage{tikz}

\newcommand{\mc}{\mathcal}

\newcommand{\dg}{\dagger}
\newcommand{\ra}{\rangle}
\newcommand{\la}{\langle}

\newcommand{\h}{\hat}

\newcommand{\beq}{\begin{equation}}
\newcommand{\eeq}{\end{equation}}
\usepackage[normalem]{ulem}
\newcommand\redsout{\bgroup\markoverwith{\textcolor{red}{\rule[0.5ex]{2pt}{0.4pt}}}\ULon}

\begin{document}

\title{Dynamical Localization Transition of String Breaking in Quantum Spin Chains}
\author{Roberto Verdel}
\email{rverdel@ictp.it}
\affiliation{Max Planck Institute for the Physics of Complex Systems, N\"othnitzer Stra{\ss}e 38,  01187 Dresden, Germany}
\affiliation{The Abdus Salam International Centre for Theoretical Physics, Strada Costiera 11, 34151 Trieste, Italy}

\author{Guo-Yi Zhu}
\email{gzhu@uni-koeln.de}
\affiliation{Max Planck Institute for the Physics of Complex Systems, N\"othnitzer Stra{\ss}e 38,  01187 Dresden, Germany}
\affiliation{Institute for Theoretical Physics, University of Cologne, Z\"{u}lpicher Straße 77, 50937 Cologne, Germany }

\author{Markus Heyl}
\email{markus.heyl@uni-a.de}
\affiliation{Max Planck Institute for the Physics of Complex Systems, N\"othnitzer Stra{\ss}e 38,  01187 Dresden, Germany}
\affiliation{Center for Electronic Correlations and Magnetism, University of Augsburg, 86135 Augsburg, Germany}

\date{\today}

\begin{abstract}

The fission of a string connecting two charges is an astounding phenomenon in confining gauge theories. 
The dynamics of this process have been studied intensively in recent years, with plenty of numerical results yielding a dichotomy:
the confining string can decay relatively fast or persist up to extremely long times. 
Here, we  put forward  a dynamical localization transition as the  mechanism underlying this dichotomy. 
To this end, we derive an effective string breaking description in the light-meson sector of a confined spin chain and show that the problem can be regarded as a dynamical localization transition in Fock space.
Fast and suppressed string breaking dynamics are  identified with delocalized and localized behavior, respectively.
We then provide a further reduction of the dynamical string breaking problem onto a quantum impurity model, 
where the string is represented as an  ``impurity'' immersed in a meson bath.
It is shown that this model features a localization-delocalization transition, giving a general and  simple physical basis to understand the qualitatively distinct string breaking regimes.
These findings are directly relevant for a wider class of confining lattice models in any dimension and could be  realized on present-day Rydberg quantum simulators. 

\end{abstract}

\maketitle

\emph{Introduction}.---The efficient implementation of gauge theories is a central target in quantum simulation~\cite{Wiese2013, Zohar_2015,  Banuls2020, 10.1088/1361-6633/ac58a4, doi:10.1098/rsta.2021.0064, doi:10.1098/rsta.2021.0062}, with  some remarkable experimental realizations achieved in recent years ~\cite{Martinez2016, Dai2017, PhysRevLett.121.030402, PhysRevA.98.032331, Kokail2019, Gorg2019, Schweizer2019, doi:10.1126/science.aaz5312, Yang2020, doi:10.1126/science.abl6277, doi:10.1126/science.abi8794, doi:10.1126/science.abi8378}.
However, the intrinsic structure of  gauge theory still poses formidable technical challenges. 
Simultaneously,  quantum spin chains, which are more amenable to quantum simulation, have been shown to be a versatile platform to emulate lattice gauge theory phenomenology. 
This has led to  recent  intensive efforts to investigate the  structure of the gauge vacuum and out-of-equilibrium transport properties under the influence of confinement in this setting~\cite{Kormos2017, PhysRevLett.122.150601, PhysRevB.99.180302, PhysRevB.99.121112, PhysRevLett.122.130603, PhysRevB.102.014308, PhysRevB.102.041118, Javier_Valencia_Tortora_2020, PhysRevB.103.L220302, Surace_2021, PhysRevB.104.L201106, Lagnese_2022, PhysRevB.105.125413, PhysRevResearch.4.L032001,  PRXQuantum.3.020316,  Birnkammer2022, PhysRevX.12.031037, PRXQuantum.3.040309, PhysRevLett.128.196601, Tan2021, Vovrosh2021, 2023arXiv230303311L}. 
Yet, various aspects of such phenomena remain to be elucidated.
In particular, numerical studies of dynamical string breaking---where a string connecting two charges decays due to  pair production~\cite{PhysRevD.71.114513, PhysRevLett.111.201601}---suggest a dichotomy for the fate of the confining string: its fission can occur relatively fast or be substantially delayed.

In this Letter, we discuss how these observations can be interpreted in terms of an \emph{underlying} dynamical localization transition.
In this picture, the \emph{localized phase} corresponds to a regime with a long-lived (prethermal) string, while the \emph{delocalized phase} to fast string breaking. 
First, we show via exact diagonalization in quantum Ising chains that two qualitatively different string dynamics are separated by a sharp threshold in the long-time behavior of dynamical quantities.
In particular, we study the survival probability and the half-chain entanglement entropy, with the former quantity serving as a direct diagnostics of string breaking.  
We then derive an effective model for the breaking of a short string by projecting onto a reduced subspace that captures resonant decay channels in the limit of vanishing transverse field.
Within this effective description, string breaking can be understood as a dynamical localization problem in Fock space.
Next, this description is  heuristically generalized  to a  quantum impurity model, where the string is effectively represented by a few-level system coupled to a meson bath.
We show that this  model  features a dynamical localization-delocalization transition, with both sides of the transition explaining the observed string breaking regimes. 
This description, independent of microscopic details, provides a general and simple physical basis to understand dynamical string breaking. 
Finally, we discuss  how  our results can be  applied to a wider class of confining lattice  models in any dimension, and potential implementations with Rydberg quantum simulators.
%


%
\emph{ String dynamics in quantum Ising chains}.---We consider the quantum  Ising model in both transverse ($h_x$) and longitudinal ($h_z$) fields, whose Hamiltonian  for $L$ spins on the ring reads
\beq
\label{eq:2.1}
\hat{H}=-J\sum_{i=1}^L \hat{\sigma}_i^z\hat{\sigma}_{i+1}^z -h_x\sum_{i=1}^L\hat{\sigma}_i^x-h_z\sum_{i=1}^L\hat{\sigma}_i^z,
\eeq
where $\hat{\sigma}^{x/z}_i$ are the Pauli matrices at site $i$, and $J>0$ is the strength of a ferromagnetic coupling. 
\begin{figure}[bt!]
	\centering
	\includegraphics[width=\columnwidth]{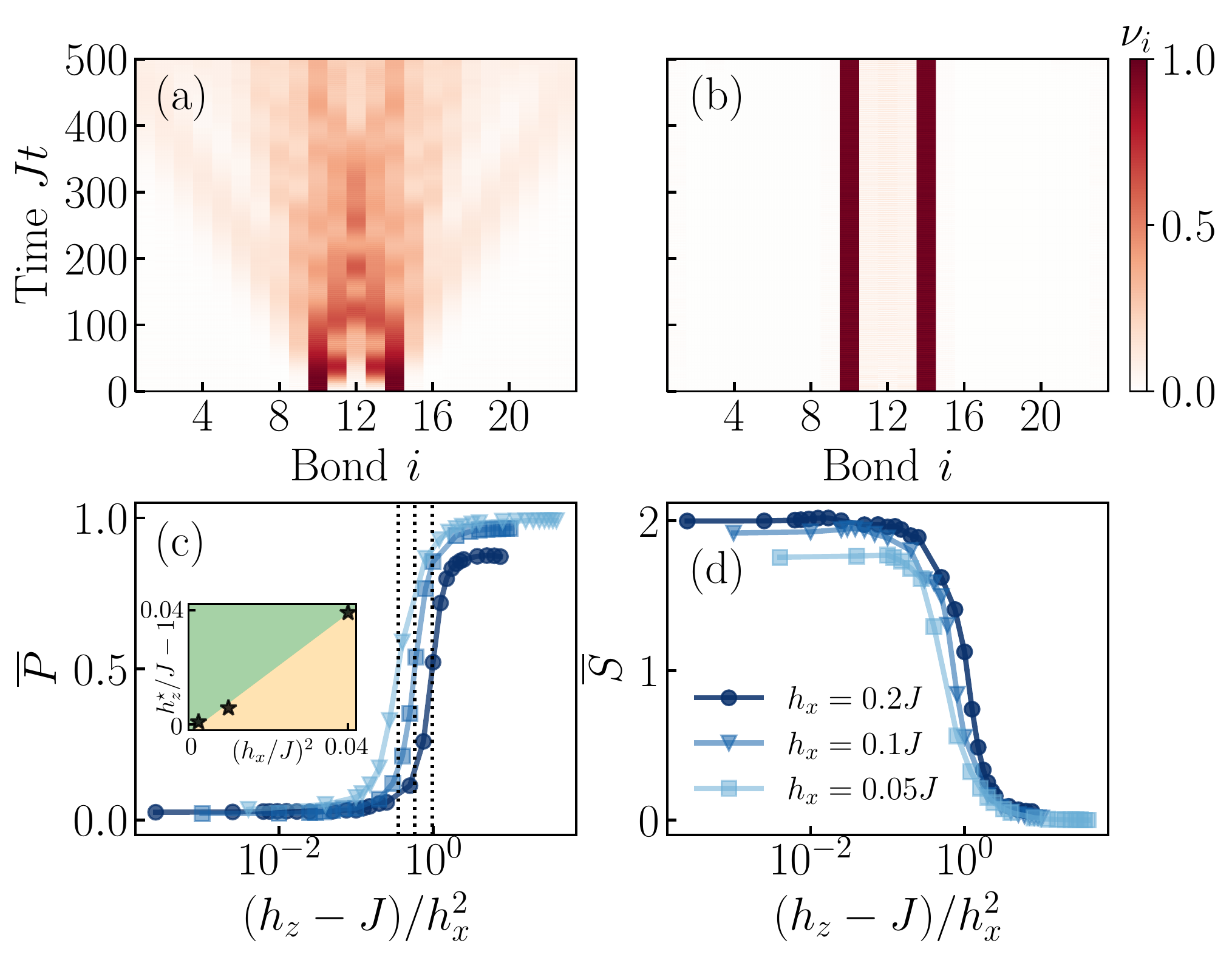}
	\caption{
	String breaking dynamics in quantum Ising chains.
        (a) [$h_z=J$] Fast and (b)  [$h_z=1.1J$] suppressed string breaking dynamics in  quantum Ising chains [Eq.~\eqref{eq:2.1}],  in terms of the spatiotemporally resolved domain wall density  $\nu_i(t)$.
        In both cases, $L=24$, $h_x=0.2J$ and  $\ell=4$ (initial string length).
        Long-time behavior of (c) the survival probability [Eq.~\eqref{eq:2.2}] and (d)  half-chain entanglement entropy [Eq.~\eqref{eq:2.3}], for various values of the magnetic fields and  $L=16, \ell=4$.
        A sharp threshold, defined by the point $h_z^\star/J$ where $\overline{P}=0.5$ [dotted lines and inset in  (c)], separates the two string breaking regimes.
        Results obtained via exact diagonalization.
	}
	\label{fig:1}
\end{figure}
The model~\eqref{eq:2.1} is of paramount  importance  in various fields---from statistical mechanics and condensed matter~\cite{McCoyWu1973, sachdev_2011} to high-energy physics~\cite{PhysRevD.18.1259, MCCOY1977219, McCoy12ising}. 
Further, it can be  naturally realized in present-day Rydberg  quantum simulators~\cite{ Bernien2017, PhysRevLett.118.063606, PhysRevX.7.041063, Lienhard2018, Guardado-Sanchez2018, PhysRevLett.120.113602,  PhysRevLett.120.113602}, and solid-state materials~\cite{Coldea2010}. 
Both integrability and $\mathbb{Z}_2$ symmetry are broken by a finite $h_z$, which induces a  
confining potential between  pairs of domain wall (DW) excitations (provided that $h_x<J$).
In this scenario, pairs of DWs form bound, mesonlike states. String breaking dynamics can then  be  probed by studying the stability of one such object under the unitary evolution generated by the Hamiltonian~\eqref{eq:2.1}.
Below, we review the main  aspects of this process 
(see also Ref.~\cite{PhysRevB.102.014308}), in the confining regime with controlled quantum fluctuations $h_x\ll J$.
A dichotomy between distinct string breaking dynamics is revealed in a simple experimentally feasible quantum quench protocol. 
The system is initially prepared in a state with an Ising electric-field string of  $\ell$ $\downarrow$-spins (in the $\sigma^z$ basis) that connects two DWs, on top of the vacuum, i.e., $|\psi_\mathrm{string}(\ell)\ra  
 \equiv|\cdots \uparrow\downarrow_{i_0}\downarrow \cdots \downarrow\downarrow_{(i_0+\ell-1)}\uparrow\cdots\ra$. 
Next, the real-time evolution of the system in Eq.~\eqref{eq:2.1} is studied at finite $h_x/J$ and $h_z/J$. 
Two qualitatively  different  dynamical  string breaking scenarios  are illustrated in Figs.~\ref{fig:1}(a) and \ref{fig:1}(b), for an initial  string of length $\ell=4$. 
The dynamics are shown in terms of the local DW density, $\nu_i(t)=\frac{1}{2}\la \h{I}-\h{\sigma}^z_i(t)\h{\sigma}^z_{i+1}(t) \ra$,
defined on the bonds between consecutive lattice sites.
In Fig.~\ref{fig:1}(a), a rapid production of new DW pairs occurs inside the string, eventually leading to its decay and emission of lighter mesons.
A subsequent proliferation of DW pairs throughout the whole chain eventually restores translation invariance, in agreement with the fact that the system~\eqref{eq:2.1} is ergodic and thermalizing at late times~\cite{PhysRevLett.111.127205}. 
This \emph{fast} 
string breaking dynamics can be understood as a consequence of underlying resonances that arise for commensurable $(J, h_z)$~\cite{PhysRevB.102.014308, c2}. 
In sharp contrast, the rapid string breaking dynamics is surprisingly absent in Fig.~\ref{fig:1}(b), up to the accessed long time $O(10^2 J^{-1})$, which is also beyond the timescale for light meson kinetics  $t \gg J/h_x^2$. 
Based on general thermalization arguments (as mentioned above), the latter regime must be  understood only as a prethermal phenomenon~\cite{Birnkammer2022}.
%


\begin{figure*}[bt!]
    \centering
    \includegraphics[width=\textwidth]{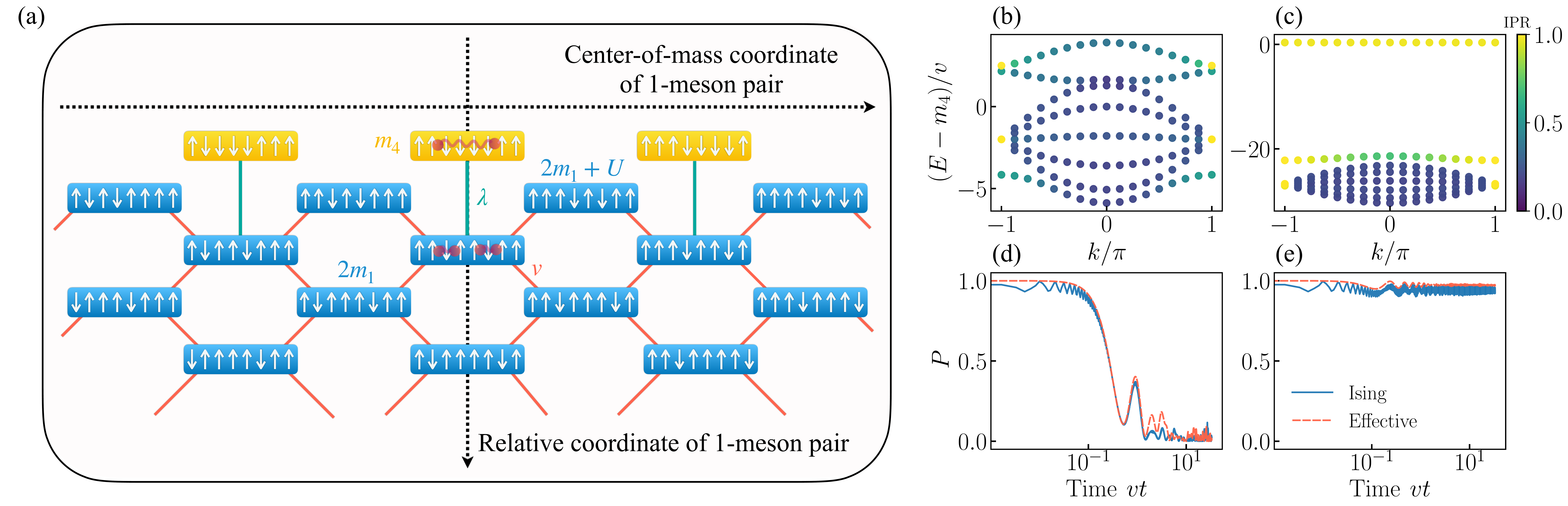}
    \caption{
    Effective graph model for string breaking dynamics.
    (a) Schematic of the effective model for a short string ($\ell=4$). Spin configurations in yellow represent string states (rest mass $m_4=12J$), whereas configurations in blue represent  1-meson pairs with energy $2m_1$ and  hopping amplitude $v=h_x^2/3J$ (red bonds).  These are the  configurations  involved, to leading order, in the resonant decay of the string (green bonds).  For illustration, confined DW pairs are depicted in some configurations as red dots joined by a wiggly line. 
    String breaking can thus be thought of as a diffusion problem in the Fock-space graph. 
    (b),(c) Energy spectrum of the effective model for $h_z=J$ and $h_z=1.02J$, respectively. In (b)  all bands have a similar energy, while in (c) there is a gap  between the ``string'' band and the continuum of 1-meson pairs. 
    Colorbar shows the IPR of  individual eigenstates [Eq.~\eqref{eq:3.2}], exhibiting a strong localization of the string modes in the latter case.
    (d),(e) Time evolution of the survival probability [Eq.~\eqref{eq:2.2}], for the respective parameters, both in the effective and full Ising model. Note the log scale in the horizontal axis and the time in unit of $v$. Parameters: $L=16, h_x=0.1J$. 
     }
    \label{fig:2}
\end{figure*}

The scenarios above have been observed in both quantum spin models~\cite{PhysRevB.102.014308, PhysRevB.99.180302, PhysRevB.102.041118} and low-dimensional lattice gauge theories~\cite{Kuhn2015, PhysRevX.6.011023, PhysRevD.98.034505, PhysRevD.99.036020,  Magnifico2020realtimedynamics, PhysRevLett.124.180602}.
However, a general picture of  how these systems cross from one regime over to the other remains to be provided. 
As a first step in this quest, we study the long-time behavior of the string survival probability 
\beq
\label{eq:2.2}
P(t) = | \la \psi_\mathrm{string}|\psi(t)\ra|^2,
\eeq
and the half-chain entanglement entropy 
\beq
\label{eq:2.3}
S(t) = S(\h{\rho}_A(t))= -\mathrm{Tr}_A[ \h{\rho}_A(t) \ln \h{\rho}_A (t)],
\eeq
where $|\psi(t)\ra$ is the time-evolved many-body wave function and $\h{\rho}_A(t)=\mathrm{Tr}_B[|\psi(t)\ra \la\psi(t) |]$ is the reduced density matrix computed on one half of the chain (cutting through the middle of the string and the opposite point on the periodic chain). 
We compute long-time averages as $\overline{\mc{O}}=\frac{1}{t_f -t_i}\int_{t_i}^{t_f}\mc{O}(t)  \mathrm{d}t$.
In our calculations we take $Jt_f=10^4$, and $Jt_i=3Jt_\mathrm{sb}$, where $Jt_\mathrm{sb} \equiv \frac{\pi}{2(h_x/J)^2}$ is a typical timescale for string breaking~\cite{PhysRevB.102.014308}. 
The long-time averages of the quantities in Eqs.~\eqref{eq:2.2} and \eqref{eq:2.3} are shown in Figs.~\ref{fig:1}(c) and  \ref{fig:1}(d), for various values of $h_z/J$ and $h_x/J$. 
We observe  a sharp threshold---defined by the point where $\overline{P}=0.5$---, which roughly scales linearly with $(h_x/J)^2$ [inset in Fig.~\ref{fig:1}(c)], and separates a regime where the string breaks ($\overline{P}\sim 0$) from one in which it persists up to the accessed timescales ($\overline{P}\sim 1$).
The behavior of $\overline{S}$  shows that  string breaking  is characterized by a significant amount of entanglement, while in   \emph{suppressed} string breaking dynamics entanglement production is strongly diminished.
\emph{String breaking as a localization problem in Fock space}.---We now derive an effective description of the above phenomenology.
Let us fix $\ell = 4$, as before.
For the considered parameter regime $h_x \ll h_z\sim J$, one can systematically project out sectors of the Hilbert space that do not participate in resonant decay channels, by applying  a Schrieffer-Wolff transformation \cite{PhysRevB.37.9753, PhysRevA.95.023621} to~\eqref{eq:2.1}, see Ref.~\cite{suppMat} for details. 
Here, the  relevant physical subspace is formed by the direct sum of the ``string'' sector and the ``1-meson pair'' sector, see Fig.~\ref{fig:2}(a). 
The former sector is spanned by the kets  $|S_j\ra= |\cdots \uparrow\uparrow\downarrow_{j}\downarrow\downarrow\downarrow\uparrow\uparrow\cdots\ra$, with a string of size $\ell=4$,  labelled by the site index $j$ of the first $\downarrow$-spin. 
The second sector comprises configurations with exactly two 1-meson particles: $|j, d\ra= |\cdots \uparrow\uparrow\downarrow_{j}\uparrow \cdots \uparrow\downarrow_{(j+d)}\uparrow\uparrow\cdots\ra$, where $2\le d\le  L/2$ ($L$ even),  is the relative distance between the two $\downarrow$-spins.
The resulting effective model reads:
\beq
\label{eq:3.1}
\h{H}_\mathrm{eff}=\h{H}_\mathrm{string}+\h{H}_\mathrm{mesons}+\h{H}_{\lambda},
\eeq
where $\h{H}_\mathrm{string}$ gives the string rest mass [$\mc{E}_s\equiv m_4=12J$]; $\h{H}_\mathrm{mesons}$ contains terms for hopping [$v=h_x^2/(3J)$], mass [$2m_1=\mc{E}_s-2v$], and  repulsive contact interaction [$U=9v/2$] of the 1-meson particles; and  $\h{H}_{\lambda}$  couples the two relevant sectors with amplitude $\lambda=-3v$; see Fig.~\ref{fig:2}(a) and Ref.~\cite{suppMat} for details.
The latter term is responsible for the processes of pair creation and recombination, and therefore, crucial for string breaking.
The energy spectrum of this model is shown in Figs.~\ref{fig:2}(b) and \ref{fig:2}(c), for different choices of parameters.
In Fig.~\ref{fig:2}(b) all  bands are close in energy, while in Fig.~\ref{fig:2}(c)  a large gap separates an isolated band (associated to string modes) from the rest.
In the latter case,  string modes are strongly localized. 
This is quantified by the colorbar in Figs.~\ref{fig:2}(b) and \ref{fig:2}(c), which shows the value of the inverse participation ratio (IPR) of individual energy eigenstates:
\beq
\label{eq:3.2}
\mathrm{IPR}(n)=\sum_{a}|\la a | n\ra|^4,
\eeq
where $\{|n\ra\}$ are eigenstates of $\h{H}_\mathrm{eff}$ and $\{|a\ra\}$ preferential basis states.
Localized behavior of $|n\ra$ occurs when $\mathrm{IPR}(n)\simeq 1$, while  $\mathrm{IPR}(n)$ vanishes as $1/D$ in the maximally delocalized case, where $D$ is the  Hilbert space dimension~\cite{PhysRevB.94.155110}. 

The evolution of the  survival probability [Eq.~\eqref{eq:2.2}], corresponding to the two cases above, is shown in Figs.~\ref{fig:2}(d) and \ref{fig:2}(e).
While in the former case, the string eventually breaks ($P \sim 0$), in the latter it survives ($P \sim 1$) up to long times.
The spectra in Figs.~\ref{fig:2}(b) and \ref{fig:2}(c) are hence identified with fast and suppressed string breaking dynamics, respectively.
String breaking can thus be seen as a dynamical localization problem in the Fock-space graph  in Fig.~\ref{fig:2}(a), where the string localizes if it is not resonantly coupled to the continuum of 1-meson pairs.
Quantitative agreement with the dynamics in the full Ising model is also observed in Figs.~\ref{fig:2}(d) and \ref{fig:2}(e), which can be systematically improved by decreasing $h_x/J$~\cite{suppMat}. 
%


%
\emph{Quantum impurity model picture}.---The above picture resembles localization phenomena in quantum impurity models (QIMs)~\cite{hewson_1993, RevModPhys.59.1, doi:10.1080/14786430500070396}. This is the basis for a further reduction of the string breaking problem.
Let us consider an elementary string breaking/fusion process: $\left(\cdots \uparrow  \uparrow \downarrow \downarrow\downarrow\downarrow\downarrow\uparrow\uparrow\cdots\right)_i \longleftrightarrow
\left(\cdots \uparrow\uparrow \downarrow\downarrow\downarrow\uparrow\downarrow\uparrow\uparrow\cdots\right)_{ii}
\longleftrightarrow \left(\cdots \uparrow\uparrow \downarrow\downarrow \downarrow\uparrow\uparrow\cdots \uparrow\downarrow\uparrow\cdots\right)_{iii} $, where a string ($i$) gets cut  near its edges via pair creation, yielding a  \emph{metastable} configuration ($ii$), and eventually, a shorter string  plus a 1-meson ($iii$)~\cite{PhysRevX.6.011023, PhysRevLett.124.180602}.
We encode the different configurations of this basic process in the internal states of a spin-1 system  (``impurity'').
Concretely, we map the symmetric and antisymmetric string states $\frac{1}{\sqrt{2}}\left[|\psi_\mathrm{string} \ra \pm |\psi_\mathrm{meta} \ra \right] $, onto the impurity states $|S^z=\pm 1\ra$, respectively, and  the state where the string has been cut and a lighter meson radiated onto $|S^z=0\ra$.
The impurity is also locally coupled to a meson bath in analogy to the picture in Fig.~\ref{fig:2}(a). 
This motivates a QIM with Hamiltonian
\begin{equation}
\label{eq:4.1}
\h{H}_\mathrm{QIM}= \h{H}_\mathrm{imp}+\h{H}_\mathrm{bath} + \h{H}_\mathrm{coup},
\end{equation} 
where $\h{H}_\mathrm{imp}= (M-\mu) (\h{S}^z)^2 + \Lambda \h{S}^z$, contains the string mass term $M$,  a chemical potential $\mu$ accounting for higher-order corrections, and a $\Lambda>0$ term, directly related  to string breaking/fusion; $\h{H}_\mathrm{bath}=\sum_{j=1}^N\big[ -T(\h{b}_j^\dg \h{b}_{j+1}+\mathrm{h.c.})+(M-2T)\h{b}^\dg_j \h{b}_j\big]$, describes a bath of light mesons  represented  by hard-core bosons with creation (annihilation) operators $\h{b}_j^\dg$ ($  \h{b}_j$), on a chain with $N$ sites,  with hopping amplitude $-T$ and maximal kinetic energy $M-2T$;  and $\h{H}_\mathrm{coup}=-T\big[ \big(1-(\h{S}^z)^2\big)\h{S}^x\h{b}_1^\dg+\mathrm{h.c.}\big]$, couples  the impurity with the bath such that if $|S^z=0\ra$ a meson  at site 1 is created, and whenever  $|S^z=\pm1\ra$ a meson at that site is annihilated~\cite{c3}.

\begin{figure}[bt!]
	\centering
	\includegraphics[width=1.\columnwidth]{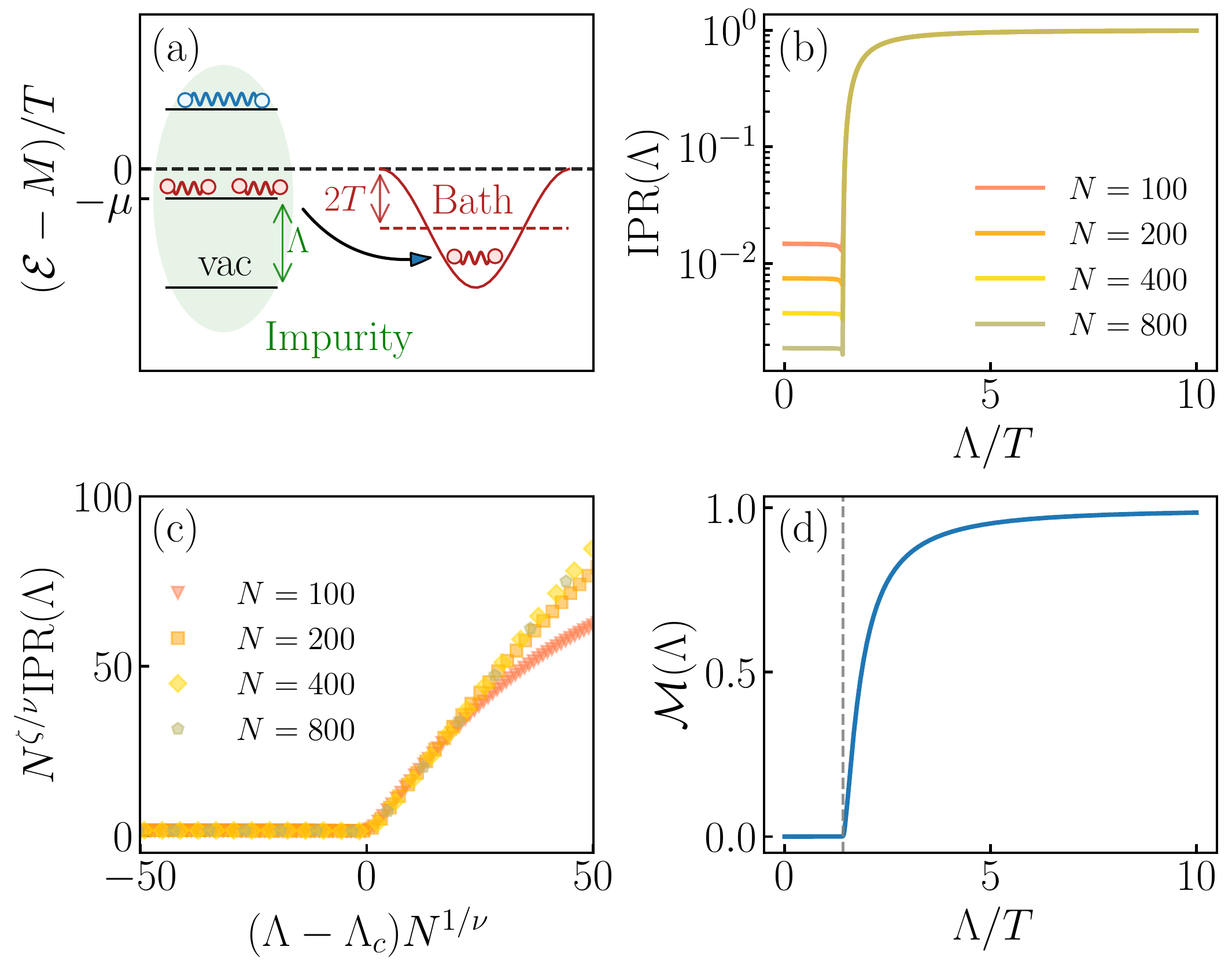}
	\caption{
	{Localization-delocalization transition in the QIM.}
	(a) Minimal string breaking/fusion as a three-level system (impurity), coupled to a bath.
    The impurity-bath coupling may or may not be resonant, yielding hybridization (as depicted here)---corresponding to string decay---or localization, respectively. 
    (b) IPR of the impurity  mode as a function of $\Lambda/T$, for various system sizes,  displaying a localization-delocalization transition. (c) Data collapse of the data in (b) using a standard finite-size scaling ansatz within the package \texttt{pyfssa}~\cite{pyfssa}, yielding $\Lambda_c/T=1.41(1)$
    and critical exponents $\zeta=1.02(5)$, $\nu=1.00(5)$.  (d) Long-time average of the spin autocorrelation function $\mc{M}(\Lambda)$, as a function of $\Lambda/T$ with $N=800$.  The dashed line indicates the transition point  $\Lambda_c/T\approx1.41$. Parameters: $M/T=10$, $\mu/T=2$. 
    }
	\label{fig:3}
\end{figure}

A schematic of this mapping is shown in Fig.~\ref{fig:3}(a) for a short string that can decay into two shorter strings. The latter can  be emitted into a meson bath, if the impurity-bath coupling is resonant, leaving the impurity in its ``vacuum'' state. Otherwise, the shorter strings can recombine back into a longer string, avoiding its decay. 
This QIM picture thus offers a distilled abstraction of the effective graph model in Fig.~\ref{fig:2}(a).
We note, however, that the mapping between these two models is not exact. Yet, as shown below, both models have significant similarities both in the behavior of their eigenstates as well as in the dynamics of the impurity and the string.

The QIM in Eq.~\eqref{eq:4.1} features a  localization-delocalization transition, explicitly shown in the single-meson limit.
In this limit the spin can be replaced by two hard-core bosons, and due to particle number conservation,  exact diagonalization is possible for large system sizes~\cite{suppMat}. 
Focusing on the IPR of individual eigenstates, we see that the impurity mode can abruptly  localize when varying  $\Lambda/T$  above a critical $\Lambda_c/T$, see Fig.~\ref{fig:3}(b), while the IPR of bulk eigenstates always vanishes (not shown). 
A standard finite-size scaling analysis~\cite{Newman}, see Fig.~\ref{fig:3}(c), yields $\Lambda_c/T=1.41(1)$ and critical exponents $\zeta=-1.02(5)$, $\nu=1.00(5)$, for the considered parameters.
Such localization-delocalization transition underlies and governs two qualitatively different spin dynamics, see 
Fig.~\ref{fig:3}(d). Here we plot  the long-time averaged spin autocorrelation function $\mc{M}(\Lambda)=\lim_{t\to \infty} \frac{1}{t} \int_0^t \mathrm{d}t' \la \h{S}^z(t')\h{S}^z(0)\ra_\Lambda$, 
where $\la \cdot \ra_\Lambda$ denotes the expectation value at a given $\Lambda/T$. 
This quantity plays an equivalent  role to the survival probability for the spin chain [Eq.~\eqref{eq:2.2}], and likewise, it vanishes on the delocalized side of the transition, while it approaches unity as we ramp up $\Lambda/T$, above the localization transition point.
Our conclusions are restricted to the lattice as we have only considered a bounded spectrum of excitations.
We expect our observations to hold beyond the limit $h_x\ll J$, as long as there exist values of $h_z/J$ for which certain decay channels lead to faster dynamics than in other regimes.
We note that our effective descriptions are valid only within the prethermal timescale of the localized regime.
Also, further localization transitions may occur around  other resonance points of the spin chain, which could involve longer decay paths~\cite{PhysRevB.102.014308, PhysRevResearch.4.L032001}, and hence, would  require to consider an impurity with more internal levels.
Regarding the Fock-space graph model, we note that adding higher-order corrections could  reshape the transition path  and change the criticality. Nevertheless, as  the effect of such higher-order terms is just a renormalization of hopping amplitudes~\cite{PhysRevB.102.041118}, we expect the physics to remain qualitatively unaltered  far from the localization transition point and deep in the two phases. 
%


%
\emph{Discussion and outlook}.---We expect our main results to be relevant for a wider class of confining theories in one and higher dimensions. 
In effect, what seems to be crucial in the applicability of the QIM picture is that  the system retains rotational symmetry, with the  radial coordinate effectively defining a one-dimensional problem,  when integrating out the rotation degree of freedom~\cite{RevModPhys.59.1}.
Fermionic bound states (e.g., baryons) could also be accounted for by changing the statistics of the bath~\cite{RevModPhys.59.1, doi:10.1080/14786430500070396}.
Finally, our observations can be experimentally realized with current quantum technologies. In particular, Rydberg atoms offer a well suited   platform, in which both the initial string states and the target unitary dynamics can be implemented in a highly controllable way~\cite{ Bernien2017, PhysRevX.7.041063, PhysRevLett.118.063606, Lienhard2018, Guardado-Sanchez2018, PhysRevLett.120.113602}.


%
\emph{Data availability}. The data shown in the figures is available on Zenodo~\cite{my_zenodo}.

\begin{acknowledgments}

\emph{Acknowledgments}. We thank T. Chanda, M. Dalmonte, P. Karpov, and M. Tsitsishvili for discussions and feedback on this work.
This project has received funding from the European Research Council (ERC) under the European Union’s Horizon 2020 research and innovation programme (grant agreement No. 853443).

\end{acknowledgments}

\bibliography{BIB}

\begin{thebibliography}{73}
\bibitem [{c2()}]{c2}%
  \BibitemOpen
  \href@noop {} {}\bibinfo {note} {The said resonances for commensurable
  $(J,h_z)$ are determined, strictly speaking, at vanishing transverse field.
  For finite $h_x\ll J$, the resonance points will be slightly shifted. This is
  illustrated, to some extent, in Fig.~\ref{fig:1}(c)}\BibitemShut {NoStop}%
\bibitem [{c3()}]{c3}%
  \BibitemOpen
  \href@noop {} {}\bibinfo {note} {Without loss of generality, we take the
  strength of the coupling between the impurity and the meson bath of the same
  order as the typical energy scale in the system, i.e. $O(T)$. See detailed
  discussion in Supp. Mat.~\cite{suppMat}}\BibitemShut {NoStop}%
\end{thebibliography}

\pagebreak
\widetext
\begin{center}
\textbf{\large Supplemental Material}
\end{center}
\setcounter{equation}{0}
\setcounter{figure}{0}
\setcounter{table}{0}
\setcounter{page}{1}
\makeatletter
\renewcommand{\theequation}{S\arabic{equation}}
\renewcommand{\thefigure}{S\arabic{figure}}
\renewcommand{\bibnumfmt}[1]{[S#1]}
\renewcommand{\citenumfont}[1]{#1}

\section{Perturbation theory for the effective model in the Fock-space graph} \label{app:perturbation_theory}

In this section, we give a detailed derivation of the effective model in Eq.~\eqref{eq:3.1}.
We restrict ourselves to the limit of weak transverse fields, $h_x\ll J$. In this limit, it is convenient to write the Hamiltonian~\eqref{eq:2.1} as

\beq
\label{eq:S.1}
\hat{H}=\hat{H}_0 + \hat{H}_1,
\eeq
where 
\begin{equation}
    \label{eq:S.2}
    \h{H}_0= -\sum_{i=1}^L \left(J \hat{\sigma}_i^z\hat{\sigma}_{i+1}^z +h_z\hat{\sigma}_i^z\right),  \,\
    \h{H}_1=-h_x\sum_{i=1}^L\hat{\sigma}_i^x,
\end{equation}
and regard $\h{H}_1$ as a small perturbation.    
Within this setting, we can perform a Schrieffer-Wolff (SW) transformation~\cite{PhysRevB.37.9753}, i.e., a unitary transformation $\mathrm{e}^{\h{S}}$ to eliminate off-diagonal terms that do not preserve $\h{H}_0$,   order by order in the strength of the perturbation.
By choosing the generator of the transformation $\h{S}=\h{S}_1+\h{S}_2+\cdots$, such that $[\h{S}_1,\h{H}_0]=-\h{H}_1$, we get the following second-order Hamiltonian~\cite{ PhysRevResearch.4.L032001, PhysRevA.95.023621}:
\begin{align}
    \label{eq:S.3}
    \h{H}_2=-\sum_{i} \Bigg\{& \Delta_+ \h{P}^{\uparrow}_{i-1}\h{\sigma}^z_i\h{P}^{\uparrow}_{i+1} + \Delta_-\h{P}^{\downarrow}_{i-1}\h{\sigma}^z_i\h{P}^{\downarrow}_{i+1} +\Delta_0\left(\h{P}^{\uparrow}_{i-1}\h{\sigma}^z_i\h{P}^{\downarrow}_{i+1}+ \h{P}^{\downarrow}_{i-1}\h{\sigma}^z_i\h{P}^{\uparrow}_{i+1}\right)\nonumber \\
    & - \left(\Delta_+-\Delta_0\right)\h{P}^{\uparrow}_{i-1}(\h{\sigma}^+_i\h{\sigma}^-_{i+1}+\mathrm{h.c.})\h{P}^{\uparrow}_{i+2} - \left(\Delta_0-\Delta_-\right)\h{P}^{\downarrow}_{i-1}(\h{\sigma}^+_i\h{\sigma}^-_{i+1}+\mathrm{h.c.})\h{P}^{\downarrow}_{i+2} \nonumber \\
    & - \left(\Delta_- -\Delta_0\right)\h{P}^{\downarrow}_{i-1}(\h{\sigma}^+_i\h{\sigma}^+_{i+1}+\mathrm{h.c.})\h{P}^{\downarrow}_{i+2}\Bigg\},
\end{align}
where $\h{P}^{\uparrow(\downarrow)}_i=(\h{1}\pm \h{\sigma}^z_i)/2$ is the projector onto $\uparrow(\downarrow)$ at site $i$,  and $\h{\sigma}^{\pm}_i=(\h{\sigma}^x_i\pm i\h{\sigma}^y_i)/2$, are spin-$\frac{1}{2}$ raising and lowering operators. We have also introduced a succinct notation for the relevant energy scales: $\Delta_0\equiv h_x^2/(2h_z)$, $\Delta_{\pm}\equiv h_x^2/(2h_z\pm 4J)$.  
The $\ell$-meson energies (with respect to the Ising vacuum energy), are obtained, to leading order, from the unperturbed Hamiltonian $\h{H}_0$ and the diagonal terms on the r.h.s. of Eq.~\eqref{eq:S.3}.  We get
\beq
\label{eq:S.4}
m_1=4J+2h_z+2(2\Delta_+-\Delta_0) +\mc{O}(h_x^4/J^3),
\eeq

\beq
\label{eq:S.5}
m_{\ell\ge 2}=4J+2\ell h_z+(\ell+2)\Delta_+ + (\ell-2)\Delta_- +\mc{O}(h_x^4/J^3).
\eeq

We now focus on the situation of interest considered in the main text.
Namely, $h_x\ll h_z=J$ and an initial string of size $\ell=4$.
The breaking of the string will necessarily yield the formation of two lighter mesons. 
Within the description above, the resonant decay channel (neglecting for a moment perturbative $(h_x/J)^2$ corrections) is into states with exactly two 1-meson particles, as it can be easily verified from Eqs.~\eqref{eq:S.4} and   \eqref{eq:S.5}, for the relevant parameters.  
Hence, to the leading order, we can project the parent Hamiltonian onto the subsector of states with a single 4-meson particle (``string'') and the subsector of two 1-meson states. 
By doing this, we  get the effective model illustrated in Fig.~\ref{fig:2} of the main text. 
Let us now explicitly write down the different terms of the this effective model. 
In the first place, we have the string and the 1-meson energies, which are given by

\beq
\label{eq:S.6}
m_1\approx 4J +2h_z -\frac{h_x^2}{3J}=6J-\frac{h_x^2}{3J}, \,\ 
\mc{E}_s \equiv m_{\ell=4}\approx 4J+8h_z=12J.
\eeq
The latter term gives  the first contribution in Eq.~\eqref{eq:3.1}. Namely,
\beq
\label{eq:S.7}
\h{H}_\mathrm{string} | S_j \ra = \mc{E}_s | S_j \ra,
\eeq
 with the same notation for the ket $| S_j \ra $ as in the main text.
On the other hand, the 1-meson energy gives the diagonal part of $\h{H}_\mathrm{mesons}$ in Eq.~\eqref{eq:3.1}, namely, $\h{H}_\mathrm{mesons}^\mathrm{diag}|j, d\ra =
2m_1  |j, d\ra$. 
We note however that this term has to be modified so as to include a hardcore repulsive interaction between 1-meson particles, which arises from the second and third terms on the r.h.s. of Eq.~\eqref{eq:S.3}, namely,  $U\equiv (2\Delta_0 - \Delta_-)=\frac{3h_x^2}{2J}$.
Next, we have to consider the off-diagonal terms on the r.h.s. of Eq.~\eqref{eq:S.3}. 
The first one of them, proportional to $\Delta_+-\Delta_0$, acts as a hopping term for the 1-meson particles. 
This energy scale, therefore, defines the kinetic energy of such particles, $v \equiv-(\Delta_+-\Delta_0)=\frac{h_x^2}{3J}$.
Hence, $\h{H}_\mathrm{mesons}$ in Eq.~\eqref{eq:3.1}, acts on a 1-meson pair ket $|j, d\ra$, as follows:
\beq
    \label{eq:S.8}
    \h{H}_\text{mesons}|j, d\ra=  (2m_1 + \delta_{d,2}U)  |j, d\ra 
    - v \Big[|j, d+1\ra+ |j+1, d-1\ra
    + |j, d-1\ra  + |j-1, d+1\ra\Big].
\eeq
The last term on the r.h.s. of  Eq.~\eqref{eq:S.3}, plays a crucial role in our description, for it describes the decay of a string of length $\ell=4$ into a 1-meson pair, with a relative distance $d=3$ and the position of the first $\downarrow$-spin in the latter configuration coinciding with that of the first $\downarrow$-spin in the former string configuration.
 This term also describes the reverse process of particle recombination into a single 4-meson particle. 
 These two process are schematically illustrated by the green bond in Fig.~\ref{fig:2}(a).
The numerical coefficient in front of this term thus sets the energy scale for the coupling $\h{H}_\lambda$, namely, $\lambda\equiv (\Delta_- - \Delta_0)=-\frac{h_x^2}{J}=-3v$.
We note that the fifth term on the r.h.s. of  Eq.~\eqref{eq:S.3}, yields no contribution when projecting onto the physically  relevant subsectors for the case at hand.
Furthermore, even though the derivation above holds strictly for $h_z=J$, it is possible to readily adjust the expressions above for $h_z \sim J$, by adding a straightforward $h_z$-dependent correction. 
Since the system is translationally invariant, it is convenient to work in Fourier space. Using $|k,d\ra =\frac{1}{\sqrt{L}} \sum_{j=1}^{L} \mathrm{e}^{-ik(j+\frac{d}{2})}|j,d\ra$, our effective model in momentum representation reads
\begin{align}
\label{eq:S.9}
\h{\mc{H}} &= \mc{E}_s \h{I} +
\sum_{k, d} (-2v + \delta_{d,2}U) |k, d\ra \la k,d|   -2v \sum_{k, d} 
 \cos\Big(\frac{k}{2}\Big) \left( |k, d\ra \la k,d+1| + |k, d\ra \la k,d-1| \right)   \nonumber \\
& -3v \sum_{k, d} \delta_{d,3} \left( |S_k\rangle \la k,d| + |k, d\ra \langle S_k | \right),
\end{align}
\noindent where  $k$ can take $L$ possible values in the Brillouin zone $-\pi<k\le \pi$, and the Kronecker delta in the last term is needed to appropriately couple  a string configuration with the relevant  state with two 1-mesons, as pointed out above. 

\begin{figure*}[bt!]
	\centering
	\includegraphics[width=1.\textwidth]{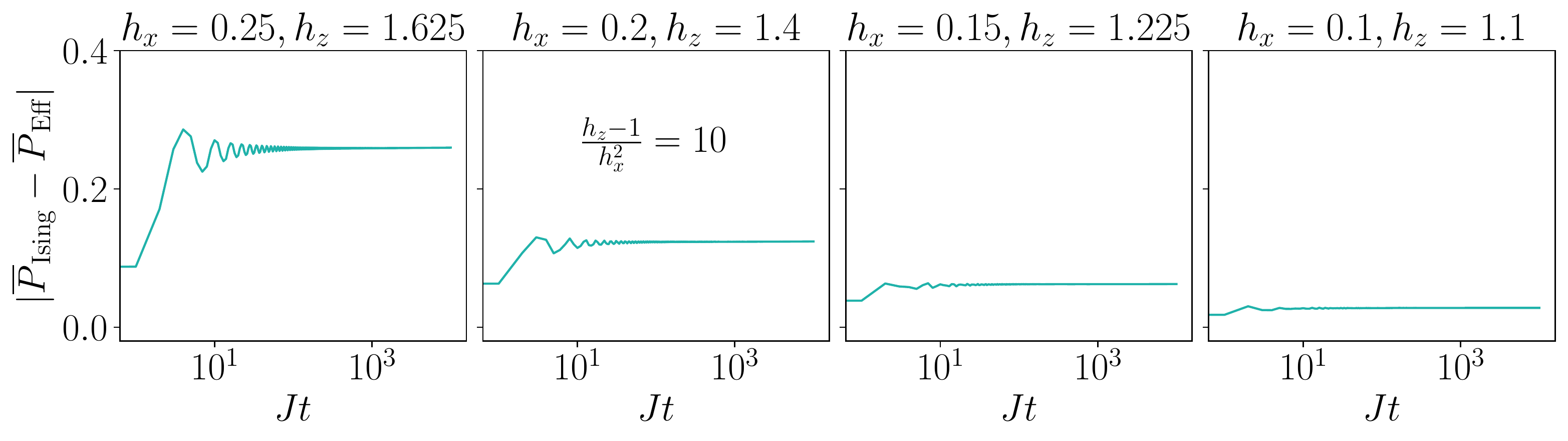}
	\caption{Convergence of the effective model in the localized regime. 
        Absolute error of the time-integrated string survival probability.
        The effective model becomes more accurate as $h_x/J$ is reduced and $h_z/J$ approaches 1,  with $\frac{h_z-1}{h_x^2}=10$ kept fixed.
        The  values of the confining field $h_z/J$ are such that they lead to localized string dynamics.
        Both $h_x$ and $h_z$ are in units of $J$.
        In all plots, we use $L=14$ and $\ell=4$.
	}
	\label{fig:S1}
\end{figure*}

While the effective model derived in this section is asymptotically exact as we let $h_x/J \to 0$ for the point $h_z=J$, it also yields accurate results for $h_z \ne J$.
This is not only illustrated in Fig.~\ref{fig:2}(e) in the main text, but also in Fig.~\ref{fig:S1}, which displays the absolute error (w.r.t. the full Ising model) of the time-integrated  string survival probability  for values of $h_z/J$ that lead to localized string dynamics.
We can see that our reduced model shows systematic convergence towards the exact dynamics in the Ising model, as we decrease $h_x/J$ and let $h_z/J \to 1$ (while keeping the ratio $\frac{h_z-1}{h_x^2}$ fixed).

Let us also note that when we apply the SW transformation to rotate the Hamiltonian in order to eliminate off-diagonal terms, we should also rotate the initial state:
\begin{equation}
\begin{split}
&|\psi_0^\mathrm{eff}\rangle = e^{\h{S}} |\psi_0\rangle = (1- h_x \sum_i\h{\sigma}_i^x + \cdots) |\cdots \uparrow \downarrow\downarrow\downarrow\downarrow \uparrow \cdots\rangle.
\end{split}
\end{equation}
Using $|\psi_0\rangle \equiv |\psi_\mathrm{string}\rangle =|\cdots \uparrow \downarrow\downarrow\downarrow\downarrow \uparrow \cdots\rangle$ as initial state for simulating the dynamics in the effective model will, instead, give us a $O(h_x/J)$-error. This error however is negligible when the survival probability is $P\sim 1$, but becomes more important when $P$ is small. Nonetheless, for the considered timescales, we still obtain a very good agreement with the exact dynamics in the full Ising model. 

\section{Exact solution of the quantum impurity model in the dilute meson limit} \label{app:free-fermion}

Here we show how to solve the quantum impurity model in Eq.~\eqref{eq:4.1}, in the single-meson limit. 
The single-meson sector is given by 
$\Big\{\lvert S^z=0 \ra \otimes \lvert b^\dg_j b_j=\delta_{j,j_0} \ra, \, \,  \lvert S^z=\pm1 \ra \otimes \lvert b^\dg_j b_j=0 \ra \Big\}$, for all $j$ and for some $1\le j_0\le N$. 
The projected Hamiltonian is thus $\mc{P}\h{H}_\mathrm{QIM}\mc{P}\equiv \h{\mc{H}}_\mathrm{QIM}$.
One can easily see that under this projection, 
the spin-1 degree of freedom representing the impurity can be replaced by two hard-core bosons with creation operators $\h{a}^\dg_\mathrm{I}$,  $\h{a}^\dg_\mathrm{II}$, with the identification $n_\mathrm{I}=1 \leftrightarrow S^z=+1$ and $n_\mathrm{II}=1 \leftrightarrow S^z=-1$ ($n_s$ being the eigenvalue of the occupation number $\h{n}_s=\h{a}_s^\dg\h{a}_s$; $s=\mathrm{I, II}$).
Therefore, we can consider a chain of $N+2$ sites, with the extra two sites  having  labels $\mathrm{I, II}$. 
This yields,
\begin{align}
    \label{eq:B.0}
\h{\mc{H}}_\mathrm{QIM}&= (M-\mu)(\h{a}_\mathrm{I}^\dg \h{a}_\mathrm{I}+\h{a}_\mathrm{II}^\dg \h{a}_\mathrm{II} ) +\Lambda(\h{a}_\mathrm{I}^\dg \h{a}_\mathrm{I}-\h{a}_\mathrm{II}^\dg \h{a}_\mathrm{II} )+\sum_{j=1}^N\Big[ -T(\h{b}_j^\dg \h{b}_{j+1}+\mathrm{h.c.})+(M-2T)\h{b}^\dg_j \h{b}_j\Big] \nonumber  \\ &-\frac{T}{\sqrt{2}}\Big[ \big(\h{a}_\mathrm{I}^\dg +\h{a}_\mathrm{II}^\dg\big)\h{b}_1+\mathrm{h.c.}\Big].
\end{align}
Furthermore, due to particle number conservation, the bilinear form above can exactly diagonalized for very large system sizes, and we can explicitly derive analytical expression for observables of interest. 
To show this, let us write this Hamiltonian in a compact way:
\begin{equation}
\label{eq:B.1}
\h{\mc{H}}_\mathrm{QIM}=\sum_{i,j} \h{c}^\dg_i H_{i,j} \h{c}_j,
\end{equation} 
with $\h{c}_i= \h{a}_i, \h{b}_i  $, and where $H$ is a Hermitian matrix (of dimension $N+2$), and hence, can be diagonalized by a unitary transformation $\mc{U}$:
\begin{equation}
\label{eq:B.2}
\h{c}_i^{\dg}=\sum_{l=1}^{N+2} \h{\alpha}_l^{\dg} (\mc{U}^{\dg})_{li}, \,\ \h{c}_j=\sum_{l=1}^{N+2} \mc{U}_{jl}\h{\alpha}_l.
\end{equation}  
Thus, $\h{H}_0$ simply reads
\begin{equation}
\label{eq:B.3}
\h{\mc{H}}_\mathrm{QIM}=\sum_{l=1}^{N+2} \epsilon_l \h{\alpha}_l^{\dg}\h{\alpha}_l.
\end{equation} 

One can readily find that the creation/annihilation operators $\h{\alpha}^\dg_l, \h{\alpha}_l$ evolve according to the following equations:
\begin{equation}
\label{eq:B.4}
\h{\alpha}_l^{\dg}(t) = \mathrm{e}^{i \epsilon_l t} \h{\alpha}^{\dg}_l, \,\ \h{\alpha}_l(t) = \mathrm{e}^{-i \epsilon_l t} \h{\alpha}_l.
\end{equation} 

The initial string state in the quantum impurity model is given by $|\psi_0\ra \equiv\frac{1}{\sqrt{2}}[|S^z=+1\ra +|S^z=-1\ra]\otimes |0\ra_\mathrm{bath}$, where $|0\ra_\mathrm{bath}$ is the vacuum state for the meson bath subsystem. In the language of the hard-core boson operators, this string state becomes:
\begin{equation}
\label{eq:B.5}
|\psi_0\ra = \frac{1}{\sqrt{2}}\Big[ \h{a}^\dg_1 |0\ra_\mathrm{imp}+ \h{a}^\dg_2 |0\ra_\mathrm{imp}\Big]\otimes |0\ra_\mathrm{bath},
\end{equation} 
where $|0\ra_\mathrm{imp}$ is the vacuum state of the impurity.

Thus, using the transformation in Eq.~\eqref{eq:B.2} and the fact that $\h{S}^z=\h{a}^\dg_1\h{a}_1-\h{a}^\dg_2\h{a}_2$, we arrive at the following expression for the spin-spin autocorrelation function:
\begin{align}
 \label{eq:B.6}
\la \psi_0| \h{S}^z(t) \h{S}^z(0) |\psi_0\ra= \frac{1}{2}&\Bigg[u^+(1,1)u^-(1,1) - u^+(1,1)u^-(1,2)+u^+(2,1)u^-(1,1)- u^+(2,1)u^-(1,2) \nonumber \\
& -u^+(1,2)u^-(2,1) +u^+(1,2)u^-(2,2) -u^+(2,2)u^-(2,1)+u^+(2,2)u^-(2,2)\Bigg],
\end{align} 
where $u^+(a,b):= \sum_l \mc{U}_{a,l} \mathrm{e}^{i\epsilon_l t} (\mc{U}^\dg)_{l,b}$ and $u^-(a,b):= \sum_m \mc{U}_{a,m} \mathrm{e}^{-i\epsilon_m t} (\mc{U}^\dg)_{m,b}$. This expression is used in the computation of the long-time average $\mc{M}(\Lambda)$.\\

\begin{figure*}[bt!]
	\centering
	\includegraphics[width=1.\textwidth]{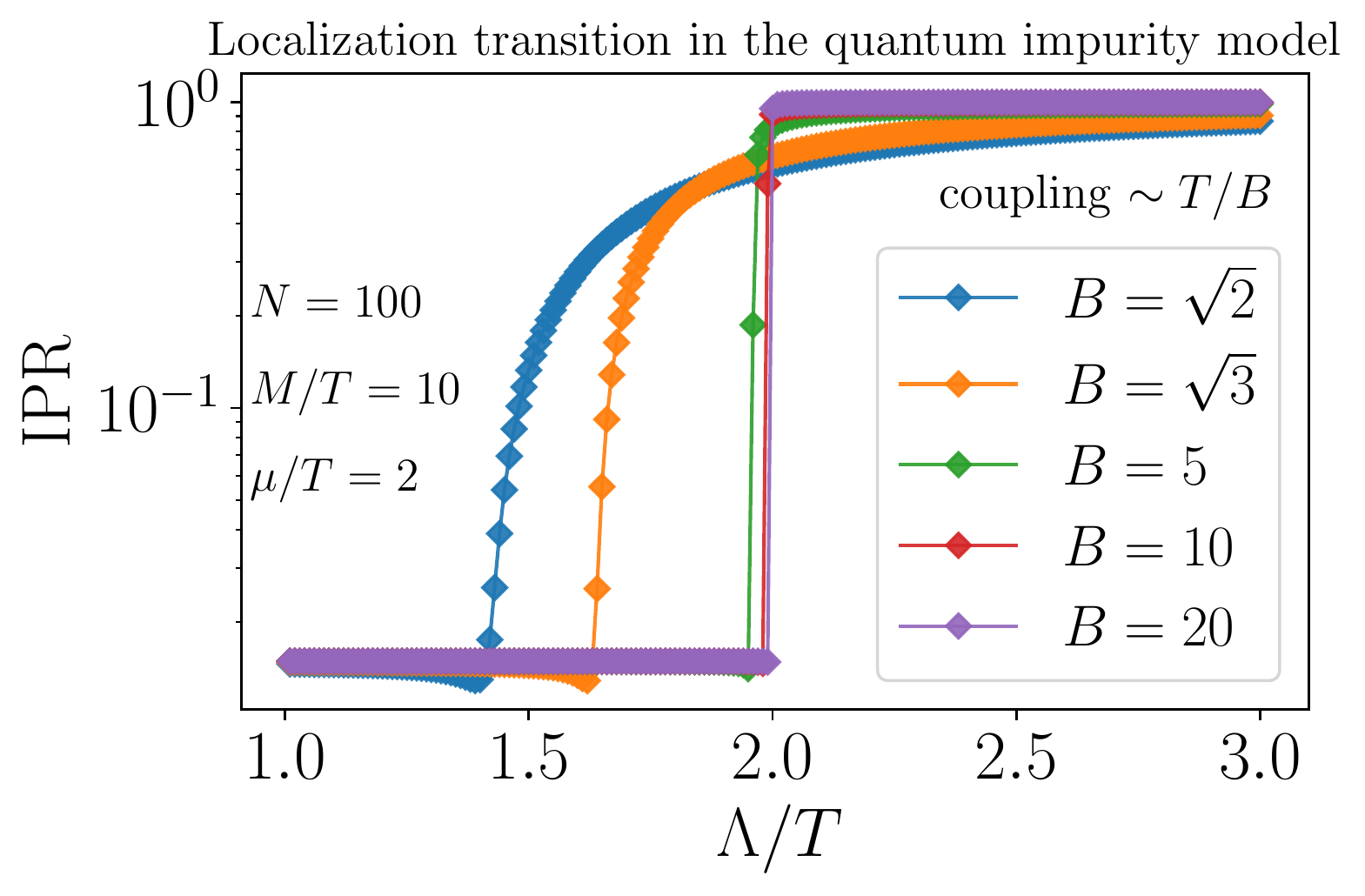}
	\caption{Localization-delocalization transition in the generalized quantum impurity model for varying strength of the coupling. IPR of the lowest-lying eigenstate of the model in Eq.~\eqref{eq:B.0} as a function of $\Lambda/T$. Here we explore the effect of adding a further free parameter $B$ modulating the strength of the coupling term in our phenomenological description [last term in Eq.~\eqref{eq:B.0}]. As the $T/B$ is reduced the critical value of $\Lambda/T$ converges to 2, as it can be readily understood from the cartoon picture in Fig.~\ref{fig:3}(a).  
	}
	\label{fig:SuppMat2}
\end{figure*}

As a final remark, we comment on how the localization-delocalization transition depends on the value of the coupling term in Eq.~\eqref{eq:B.0}.
This is illustrated in Fig.~\ref{fig:SuppMat2}, where we introduce a new parameter $B$ modulating the strength of the said term. 
As the coupling between the impurity and the bath is decreased, we observe that the critical value of $\Lambda/T$ approaches 2 (for the specific parameters used here, in particular, $\mu/T=2$). 
Further, the transition becomes sharper and sharper. 
This can be easily understood from the cartoon picture drawn in Fig.~\ref{fig:3}(a), which precisely illustrates the limiting case of $B/T\to \infty$. 
At any rate, the overall qualitative result remains unaltered.

\end{document}